\documentclass[
 aip,
 amsmath,amssymb,
 reprint,
 ]{revtex4-1}

\usepackage{graphicx}
\usepackage{dcolumn}
\usepackage{bm}

\usepackage[utf8]{inputenc}
\usepackage[T1]{fontenc}
\usepackage{mathptmx}
\usepackage{etoolbox}
\usepackage{color}

\makeatletter
\def\@email#1#2{%
 \endgroup
 \patchcmd{\titleblock@produce}
  {\frontmatter@RRAPformat}
  {\frontmatter@RRAPformat{\produce@RRAP{*#1\href{mailto:#2}{#2}}}\frontmatter@RRAPformat}
  {}{}
}
\makeatother
\begin{document}

\preprint{AIP/123-QED}

\title[Josephson diode effect in a ballistic single-channel nanowires]{Josephson diode effect in a ballistic single-channel nanowire}
\author{Julia~S.~Meyer}
\author{Manuel~Houzet}
 \email{julia.meyer@univ-grenoble-alpes.fr.}
\affiliation{Univ.~Grenoble Alpes, CEA, Grenoble INP, IRIG, PHELIQS, 38000 Grenoble, France
}

\date{\today}

\begin{abstract}
When time-reversal and inversion symmetry are broken, superconducting circuits may exhibit a so-called diode effect, where the critical currents for opposite directions of the current flow differ. In recent years, this effect has been observed in a multitude of systems and the different physical ingredients that may yield such an effect are well understood. On a microscopic level, the interplay between spin-orbit coupling and a Zeeman field may give rise to a diode effect in a single Josephson junction. However, so far there is no analytical description of the effect within a simple model. Here we study a single channel nanowire with Rashba spin-orbit coupling and in the presence of a Zeeman field. We show that the different Fermi velocities and spin projections of the two pseudo-spin bands lead to a diode effect. Simple analytical expressions for the diode efficiency can be obtained in limiting cases.
\end{abstract}

\maketitle

\begin{quotation}
The so-called superconducting diode effect has attracted a lot of attention recently~\cite{review,review2}. In systems breaking inversion and time-reversal symmetry, the critical currents in opposite directions may be different. Experimentally this has been observed in a variety of bulk superconductors~\cite{exp-bulk,bulk_p1,bulk_p2,bulk_p3,bulk_p4}, as well as in circuits containing Josephson junctions~\cite{exp-old,exp-JJ1,exp-JJ2,JJ_p1,JJ_p2,JJ_p3,JJ_p4,sergey}. One of the possible origins of the diode effect in Josephson junctions is the interplay between Rashba spin-orbit coupling and an external magnetic field~\cite{multichannel1,multichannel1-prb,multichannel2}. The diode effect is then linked to the so-called $\phi_0$-junction behavior~\cite{JA1,JA2,JA3,multichannel1,multichannel1-prb}, where the current phase-relation acquires a field-tunable phase shift and an anomalous supercurrent flows in the absence of a phase bias. The simplest current-phase relation yielding these effects is $I=I_{c1}\sin(\phi-\phi_{01})$. By contrast, the diode effect requires a current-phase relation with more than one harmonic. The simplest case is provided by the presence of a second harmonic, $I_{c2}\sin(2\phi-\phi_{02})$, with a phase shift $\phi_{02}\neq2\phi_{01}$. While such a phenomenological current-phase relation has been used in the literature~\cite{exp-JJ2,Hu,etamax}, a microscopic derivation  has been missing. Here we show that, in ballistic single-channel Josephson junctions, such a current-phase relation can indeed be realized, both in short and long junctions at finite temperature, thus allowing one to characterize their diode efficiency. In this paper, we will concentrate on the cases where analytical results are available.
\end{quotation}


	\begin{figure}
		\includegraphics[width=0.7\columnwidth]{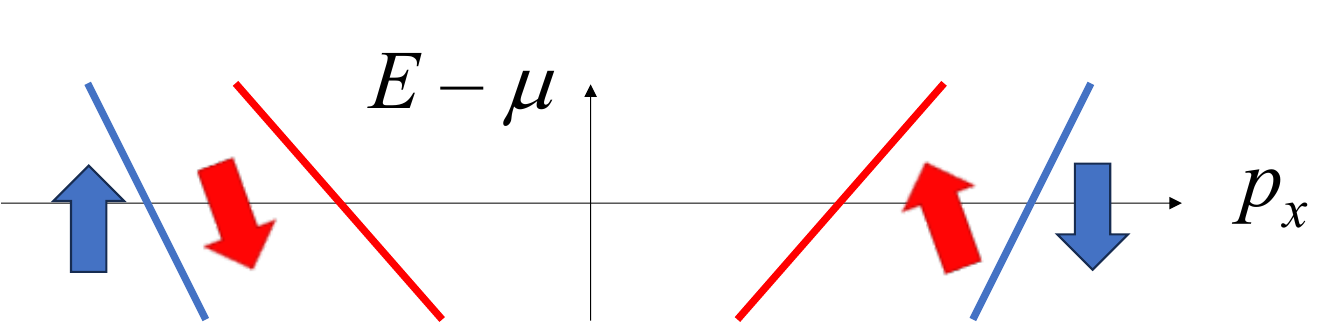}
		\caption{\label{fig-dispersion} Linearized band structure  in the vicinity of the Fermi level. The $j=+$ band shown in blue and the $j=-$ band shown in red have different Fermi velocities as well as different spin projections.}
	\end{figure}

We consider a simple 2D model of a nanowire with Rashba spin-orbit coupling~\cite{NW-1,NW-last}  in the presence of an external magnetic field. Due to the Rashba spin-orbit coupling, an electric field perpendicular to the plane generates an effective spin-orbit magnetic field in the plane and perpendicular to the nanowire. A diode effect may arise if the external magnetic field is aligned with the spin-orbit field.

A  single channel with different Fermi velocities, $v_j=\bar v+j\delta v/2$, and Zeeman energies, $h_j=\bar h+j\delta h/2$, resulting from different spin projections, for the two pseudo-spin bands $j=\pm$, see Fig.~\ref{fig-dispersion},  requires at least one of two ingredients (see Supplementary Material):  
a finite transverse width such that the spin-orbit coupling associated with the transverse motion comes into play or a magnetic field $\vec h_\perp$ transverse to the spin-orbit field.
In the former case, $\delta v/\bar v \sim E_{so}^{5/2}\sqrt{\bar\mu}/\omega_0^3$, whereas $\delta h/\bar h \sim E_{so}^{3/2}\sqrt{\bar\mu}/\omega_0^2$ is typically larger. Here $E_{so}=m\alpha^2/2$ is the spin-orbit energy associated with the Rashba coupling $\alpha$ and $m$ is the effective electron mass, while $\bar\mu$ is an effective chemical potential and $\omega_0$ is the energy scale associated with the transverse confining potential.
In the latter case,  the difference in Fermi velocities and effective fields is determined by $\delta v/\bar v,\delta h/\bar h \sim h_\perp^2/(\sqrt{E_{so}}\bar\mu^{3/2})$. The field $\vec h_\perp$ may be chosen along the nanowire, which is the direction considered in order to realize topological superconductivity with Majorana end modes in proximitized nanowires~\cite{majorana1,majorana2}. When superconductivity is induced, $h_\perp$ also modifies the induced gaps $\Delta_j=\bar \Delta+j\delta \Delta/2$ in such a way that $\Delta_-/\Delta_+=h_-/h_+$~\cite{nesterov}. 

 We neglect the orbital effect, which is expected to be small as long as  $h\ll g\mu_B \Phi_0/(ad)$, where $g$ is the $g$-factor, $\mu_B$ the Bohr magneton, $\Phi_0$ the flux quantum, and $a$ is the extension of the wavefunction in the direction perpendicular to the plane. Furthermore, the relevant length $d$ is set by the smaller of the coherence length $\xi=\bar v/\bar\Delta$ or the length of the junction $L$. As the enhancement of the $g$-factor induced by spin-orbit coupling compensates an associated reduction of the effective mass, the above condition at the field that optimizes the diode effect  $h\sim \Delta$ for the effect due to the reservoirs and $h\sim \bar v/L$ for the effect due to the normal part of the junction, see below) reduces to $a\ll \lambda_F$, where $\lambda_F=2\pi/(m\bar v)$ is the Fermi wavelength; it is fulfilled in a semiconducting nanowire at low density~\cite{note2D-3D}. (We use units where $\hbar=k_B=1$.)

Linearizing the spectrum close to the Fermi level, one thus obtains an effective Hamiltonian
\begin{eqnarray}
H_{\rm eff}^{\rm N}=\sum_{j=\pm}(j v_j p_x-h_j)\sigma_z-\bar\mu,
\end{eqnarray}
where $p_x$ is the longitudinal momentum and $\sigma_z$ is a Pauli matrix in pseudo-spin space.
Adding superconductivity to describe a Josephson junction, such that the normal part of the junction extends from $-L/2$ to $L/2$, one finds
\begin{eqnarray}
\label{eq:Heff}
H_{\rm eff}=\sum_j\left[(jv_j p_x\sigma_z-\mu)\tau_z- h_j\sigma_z+\Delta_j\tau_++\Delta_j^*\tau_-\right],
\end{eqnarray}
where $\tau_i$ ($i=x,y,z$) are Pauli matrices in particle-hole space  and $\tau_\pm=\tau_x\pm i\tau_y$. The spatial dependence of the Zeeman field and order parameter are given as $h_j(x)=h_j^{\rm N}\theta(L/2-|x|)+h_j^{\rm S}\theta(|x|-L/2)$ and $\Delta_j(x)=\Delta_j\exp[i\,{\rm sign}(x)\phi/2]\theta(|x|-L/2)$, where $\Delta_j$ and $\phi$ are the superconducting gaps of the two pseudo-spin bands and phase difference, respectively.~\cite{footnote-gaps} Here the magnetic field $h_j^{\rm N}$ in the junction and the magnetic field $h_j^{\rm S}$ in the superconducting leads may be different due to magnetic screening. 
Note that an homogeneous version of Eq.~\eqref{eq:Heff} may be used to analyze the \lq\lq bulk\rq\rq\ diode efficiency associated with the asymmetry of depairing currents in a 1D model of bulk superconductor\cite{bulk-theo,vs-Dolcini}.

Both fields $h_j^{\rm N}$ and $h_j^{\rm S}$ contribute to the Josephson diode effect~\cite{dolcini,nesterov,tanaka}. We obtain the supercurrent
using Eq.~(3) in Ref.~\onlinecite{dolcini} for a ballistic Josephson junction and summing over both pseudo-spin bands: 
\begin{eqnarray}
I_J&=&-4eT\frac d{d\phi}\Re\!\!\!\sum_{j=\pm;\nu\geq0}\!\!\!\ln\left[1\!+\!a_j^2(\omega_\nu)e^{-2\omega_\nu/E_{Lj}}e^{i(\phi+j\phi_j)}\right]/\nonumber\\
\label{eq-current-general}
\end{eqnarray}
Furthermore, $E_{Lj}=v_j/L$ and $\phi_j=2h_j^{\rm N}/E_{Lj}$ are Thouless energies and phase shifts associated with electron propagation in the normal region.  
$a_\pm(\omega_\nu)=(\omega_\nu\mp ih_\pm^{\rm S})/\Delta_\pm-[{(\omega_\nu\mp ih_\pm^{\rm S})^2/\Delta_\pm^2+1}]^{1/2}$ with the fermionic Matsubara frequencies $\omega_\nu=2\pi T(\nu+1/2)$ at temperature $T$ accounts for scattering at the normal/superconductor interfaces. (Restituting $k_B$ and $\hbar$, the prefactor yielding the current scale is given as $eT\to ek_BT/\hbar$.)

The critical currents for opposite current directions are defined as $I_c^>={\rm max}_\phi I_J(\phi)$ and $I_c^<=-{\rm min}_\phi I_J(\phi)$. Then Eq.~\eqref{eq-current-general} allows one to obtain the diode efficiency
$ \eta=(I_c^>-I_c^<)/(I_c^>+I_c^<)$ for arbitrary fields, length of the junction, and temperature.


It turns out the diode effect due to the field in the junction behaves quite differently than the one due to the field in the leads. We will start by discussing the effect of the field in the junction only. To do so, we will assume that the magnetic field is confined to the junction. 
Then Eq.~\eqref{eq-current-general} simplifies to
\begin{eqnarray}
I_J&=&\sum_{j=\pm;k>0} I_{kj}\sin(k(\phi+j\phi_j))
\label{eq-current}
\end{eqnarray}
with\begin{equation}
I_{kj}=-4eT\sum_{\nu\geq0} (-1)^kka^{2k}(\omega_v)e^{-2k\omega_\nu/E_{Lj}},
\end{equation}
Here $a(\omega_\nu)=\omega_\nu/\Delta-[{\omega_\nu^2/\Delta^2+1}]^{1/2}$, where $\Delta$ is the common gap for both pseudo-spin bands.


In general, all harmonics contribute to the current-phase relation of a ballistic Josephson junction. Typically, the higher the harmonics, the stronger their suppression with temperature. When the two first harmonics dominate over the other ones, the current-phase relation \eqref{eq-current} can be rewritten in the form\cite{footnote-squid}
\begin{eqnarray}
I_J=I_{c1}\sin(\phi-\phi_0)+I_{c2}\sin(2(\phi-\phi_{0})+\delta_{12})
\end{eqnarray}
with
\begin{eqnarray}
I_{ck}&=&\sqrt{{I_{k+}}^2+{I_{k-}}^2+2{I_{k+}}{I_{k-}}\cos(k(\phi_++\phi_-))},\\
\phi_{0k}&=&-\arctan\frac{I_{k+}\sin(k\phi_+)-I_{k-}\sin(k\phi_-)}{I_{k+}\cos(k\phi_+)+I_{k-}\cos(k\phi_-)},
\end{eqnarray}
such that $\phi_0=\phi_{01}$ and $\delta_{12}=2\phi_{01}-\phi_{02}$.  The diode efficiency is non-vanishing unless $I_{c1}=0$ or $I_{c2}=0$ or $\delta_{12}=0\,{\rm mod}\,\pi$. The maximal diode efficiency~\cite{etamax} $\eta_{\rm max}=1/3$ is achieved at $I_{c2}=I_{c1}/2$ and $\delta_{12}=\pi/2\;{\rm mod}\;\pi$. 

Typically $I_{c2}\ll I_{c1}$. Then the diode efficiency takes the simple form
\begin{eqnarray}
\eta\approx-\frac{I_{c2}}{I_{c1}}\sin\delta_{12}.
\label{eq-diode12}
\end{eqnarray}
The result can be further simplified introducing $I_{kj}=\bar I_k+j\delta I_k/2$ and $\phi_j=\bar \phi+j\delta \phi/2$. At weak spin-orbit coupling, $\delta v/\bar v,\delta h/\bar h\ll1$ yields $\delta I_k\ll \bar I_k$ and $\delta \phi\ll \bar \phi$, such that
$I_{ck}\approx2\bar I_k\left|\cos(k\bar\phi)\right|$
and 
\begin{equation}
\phi_{0k}\approx-\left(\frac{k\delta\phi}2+\frac{\delta I_k}{2\bar I_k}\tan(k\bar\phi)\right).
\end{equation}
Note that $\bar\phi=2\bar h/\bar E_L$ with $\bar E_L=\bar v/L$ and $\delta\phi=(\delta h/\bar h-\delta v/\bar v)\bar\phi/2$. If $\bar\phi\ll1$, the diode efficiency is linear in the magnetic field and given as
\begin{eqnarray}
\eta\approx\frac{\bar I_2}{\bar I_1}\left(\frac{\delta I_1}{\bar I_1}-\frac{\delta I_2}{\bar I_2}\right)\bar\phi.
\end{eqnarray}
 If $\delta I_2/\bar I_2=\delta I_1/\bar I_1$, a residual effect $\propto \bar h^3$ remains, namely
$\eta\approx(\delta I_1/\bar I_1)\bar\phi^3$. As $\delta_{12}$ increases with $\bar h$, the diode efficiency changes sign at larger fields, when $\delta_{12}$ exceeds $\pi$, according to Eq.~\eqref{eq-diode12}.

As we discuss below, a current-phase relation with essentially two harmonics may be realized in a junction with arbitrary length at $T_c-T\ll T_c$, where $T_c$ is the critical temperature of the leads. In a long junction, lower temperatures $E_{Lj}\ll T\ll T_c $ are sufficient.


Near $T_c$, the gap $\Delta$ vanishes. 
Thus $\Delta\ll\omega_\nu$, such that $a(\omega_\nu)\approx-\Delta/(2\omega_\nu)$. As a consequence, $I_{kj}\propto \Delta^{2k}$ and we may keep $k=1,2$ in Eq.~\eqref{eq-current} only with 
\begin{eqnarray}
I_{kj}=-4eT_c(-1)^kk\sum_\nu(\Delta/2\omega_\nu)^{2k}
e^{-2k\omega_\nu/E_{Lj}}.
\end{eqnarray}
In the limit $L\to0$ or $E_{Lj}\to \infty$, both pseudo-spin bands yield supercurrents of the same magnitude, i.e., $\delta I_k=0$. Thus, the finite length has to be taken into account in order to obtain a diode effect.  In a short junction, with logarithmic accuracy, this yields 
\begin{eqnarray}
I_{1j}&\approx&\frac{\pi^2}{2}eT_c\left(\frac{\Delta}{2\pi T_c}\right)^2\left(1-\frac4{\pi^2}\frac{2\pi T_c}{E_{Lj}}\ln\frac{E_{Lj}}{2\pi T_c}\right),\\
I_{2j}&\approx&-\frac{\pi^2}{24}eT_c\left(\frac{\Delta}{2\pi T_c}\right)^4.
\end{eqnarray}
Thus $I_{c2}/I_{c1}\approx-\left(\Delta/2\pi T_c\right)^2/12$ and
\begin{eqnarray}
\delta_{12}\approx-\frac{4}{\pi^2}\frac{2\pi T_c}{\bar E_L}\ln\left(\frac{\bar E_L}{2\pi T_c}\right)\frac{\delta v}{\bar v}\frac{\bar h}{\bar E_L} \ll 1.
\end{eqnarray}
The diode efficiency $\eta_{\rm short}$ is then given by Eq.~\eqref{eq-diode12}.


In a long junction, the contributions from Matsubara frequencies $\omega_{\nu>0}$ are suppressed when $E_{Lj}\ll T\ll T_c$ and we may approximate $a(\omega_0)\approx-1$. As a consequence, $I_{kj}\propto \exp[-k2\pi T/E_{Lj}]$  and we may keep $k=1,2$ in Eq.~\eqref{eq-current} only with 
\begin{eqnarray}
I_{kj}&\approx&-4eT(-1)^kke^{-k2\pi T/E_{Lj}}.
\end{eqnarray}
At small $\bar h$, this yields
\begin{equation}
\delta_{12}\approx\left(\tanh\frac{2\pi T\delta v}{\bar E_L\bar v}-\tanh\frac{\pi T\delta v}{\bar E_L\bar v}\right)\frac{2\bar h}{\bar E_L}.
\end{equation}
At $\delta v/\bar v\ll \bar E_L/T$, the result simplifies to $\delta_{12}\approx-2\pi (T\bar h/\bar E_L^2)\delta v /\bar v \ll 1$. The corresponding diode efficiency is exponentially suppressed, $\eta_{\rm long}\propto\exp[-2\pi T/\bar E_L]$.

 
Though the current-phase relation of a ballistic Josephson junction at  $T=0$  contains many harmonics and displays jumps, analytical results can be obtained both for short and long junctions.

 	\begin{figure}
		(a)\includegraphics[width=0.45\columnwidth]{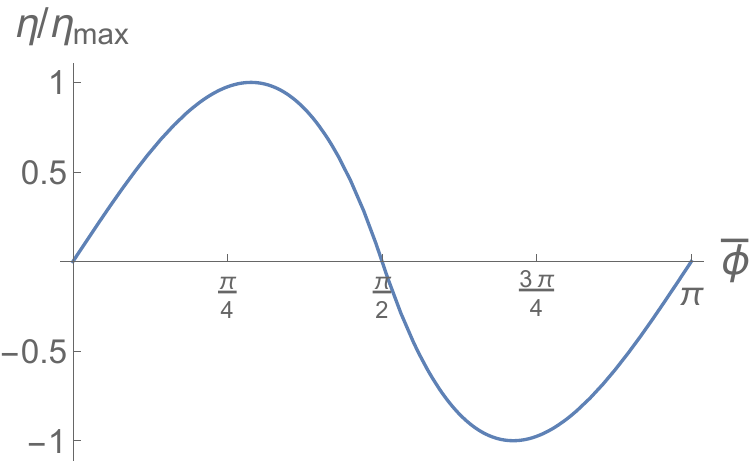}\hfill(b)\includegraphics[width=0.45\columnwidth]{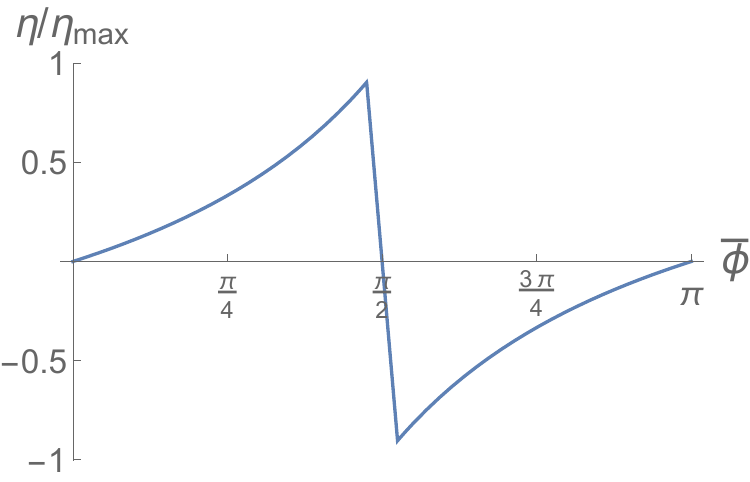}
		\caption{\label{fig-diode-T0} The diode efficiency at $T=0$ due to the magnetic field in the junction as a function of the phase shift $\bar\phi \propto \bar h$ for (a) a short junction and (b) a long junction. In the latter case, the width of the phase interval around $\pi/2$ where $\eta$ decreases with $\bar\phi$ is set by $\delta v/\bar v$.}
	\end{figure}


To obtain a diode effect in a short junction, we need to keep both the phase acquired in the junction ($\propto h_jL$) and the leading correction in $\Delta/E_{Lj}\ll 1$.
While the main contribution to the Josephson current is carried by Andreev bound states with energy $E_{{\rm A}j}=\Delta\cos((\phi+j\phi_j)/2)$, the correction comes from the continuum~\cite{Levchenko2006}, which carries a current in the opposite direction. With logarithmic accuracy, one finds
\begin{eqnarray}
I_J
&=&\frac{e\Delta}2\sum_{j=\pm}\sin\frac{\phi+j\phi_j}2{\rm sign}\left(\cos\frac{\phi+j\phi_j}2\right)\quad\nonumber\\
&&\qquad\times\left[1- \frac4\pi\left(\frac\Delta{E_{Lj}}\ln\frac{E_{Lj}}\Delta\right)\left|\cos\frac{\phi+j\phi_j}2\right|\right].
\end{eqnarray}
Thus, we find an average critical current
$\bar I_c=(e\Delta/2)\left(1+\left|\cos\bar\phi\right|\right)$
and a diode efficiency
\begin{eqnarray}
\eta_{0\,{\rm short}}&=&\frac2\pi\left(\frac\Delta{\bar E_L}\ln\frac{\bar E_L}\Delta\right)\frac{\delta v}{\bar v}\frac{\sin(2\bar\phi)}{1+|\cos\bar\phi|}.
\end{eqnarray}
The maximal diode efficiency is $\eta_{0\,{\rm short}}^{\rm max}=\sqrt{10\sqrt{5}-22}\,(2\Delta\delta v/\pi\bar E_L\bar v)\ln(\bar E_L/\Delta)$. The dependence on $\bar\phi$ is shown in Fig.~\ref{fig-diode-T0}(a).

 
 In a long Josephson junction, the current-phase relation at $T=0$ of each pseudo-spin band has a sawtooth shape~\cite{Ishii1970}. The total current reads
\begin{equation}
 I_J=\sum_{j=\pm} eE_{Lj}\Bigg[\frac{\phi+j\phi_j}{2\pi} 
 -{\rm Int}\left(\frac{\phi+j\phi_j+\pi\,{\rm sign}(\phi+j\phi_j)}{2\pi}\right)\Bigg].\;
 \end{equation}
Defining $x=\bar\phi/\pi- {\rm Int}(\bar\phi/\pi)$, this yields an average critical current
\begin{eqnarray}
 \bar I_c=e\bar E_L\begin{cases}
1-x,& x<\frac12({1-\frac{\delta v}{2\bar v}}),\\
\frac12({1+\frac{\delta v}{2\bar v}}),& \frac12({1-\frac{\delta v}{2\bar v}})<x<\frac12({1+\frac{\delta v}{2\bar v}}),\\
x, &\frac12({1+\frac{\delta v}{2\bar v}})<x,\\
\end{cases}
\end{eqnarray}
and a diode efficiency
\begin{eqnarray}
\eta_{0\,{\rm long}}
&\approx&\begin{cases}
\frac{\delta v}{2\bar v}\frac{x}{1-x},& x<\frac12({1-\frac{\delta v}{2\bar v}}),\\
1-2x,&\frac12({1-\frac{\delta v}{2\bar v}})<x<\frac12({1+\frac{\delta v}{2\bar v}}),\\
-\frac{\delta v}{2\bar v}\frac{1-x}{x}, &\frac12({1+\frac{\delta v}{2\bar v}})<x.\\
\end{cases}
\end{eqnarray}
The maximal diode efficiency is $\eta_{0\,{\rm long}}^{\rm max}=\delta v/(2\bar v)\gg\eta_{0\,{\rm short}}^{\rm max}$. I.e., the diode efficiency is enhanced in a long junction. The dependence on $\bar\phi$ is shown in Fig.~\ref{fig-diode-T0}(b).


We finally turn to the effect of the magnetic field in the leads. As shown in Ref.~\onlinecite{dolcini}, the magnetic field $h_j^{\rm S}$ yields a diode effect in a topological junction where only one pseudo-spin band is present. It is due to the fact that the magnetic field induces a current in the leads, which are formed of helical edges ``proximitized'' by a conventional superconductor. Indeed, if superconductivity in the edges were intrinsic, the field would induce a spatially modulated order parameter corresponding to a zero-current state. By contrast, the uniform order parameter imposed by the conventional superconductor covering the edges forces a current along the edges and, thus, across the junction. In the short junction limit, $L\to 0$, and $h\equiv h_j^{\rm S} <\Delta$, the current-phase relation of a single pseudo-spin band at $T=0$ reads~\cite{dolcini,footnote-fu}
\begin{equation}
I_j(\phi)=\frac{eh}\pi+\frac{e\Delta}2\sin\frac\phi2{\rm sign}\left(\cos\frac{\phi+j\phi_j}2\right),
\end{equation}
where $\phi_j=2\arcsin(h_j^{\rm S}/\Delta)$.
The diode efficiency is given as 
\begin{equation}
\eta_{\rm topo}^{\rm S}=\frac{{4h}/{\pi\Delta}-1+\sqrt{1-{h^2}/{\Delta^2}}}{1+\sqrt{1-{h^2}/{\Delta^2}}}.
\end{equation}
Its maximum value $\eta_{\rm topo}^{\rm max}=4/\pi^2$ is achieved at $h/\Delta=4\pi/(\pi^2+4)$.
In the presence of both pseudo-spin bands, there is a partial cancellation between the bands as the currents they carry in the leads are opposed~\cite{nesterov}. 
The residual diode efficiency is given as~\cite{footnote-topo}
\begin{equation}
\eta^{\rm S}= \begin{cases}\frac{\frac{2\delta h}{\pi\bar\Delta}-\sqrt{1-{h_-^2}/{\Delta_-^2}}+\sqrt{1-{h_+^2}/{\Delta_+^2}}}{\sqrt{1-{h_-^2}/{\Delta_-^2}}+\sqrt{1-{h_+^2}/{\Delta_+^2}}}&\frac {h_-}{\Delta_-}<\sqrt{1-\frac{\delta\Delta^2}{4\bar\Delta^2}}\\
\frac{\frac{2\delta h}{\pi\bar\Delta}-\delta\Delta/(2\bar\Delta)+\sqrt{1-{h_+^2}/{\Delta_+^2}}}{\delta\Delta/(2\bar\Delta)+\sqrt{1-{h_+^2}/{\Delta_+^2}}}&\sqrt{1-\frac{\delta\Delta^2}{4\bar\Delta^2}}<\frac {h_-}{\Delta_-}<1\end{cases}
\end{equation}
Using $h_+/\Delta_+=h_-/\Delta_-\equiv h/\Delta$, 
we find a maximal diode efficiency $\eta^{\rm S}_{\rm max}=(2/\pi)\sqrt{1-{\delta\Delta^2}/({4\bar\Delta^2})}$ at $h/\Delta=\sqrt{1-{\delta\Delta^2}/({4\bar\Delta^2})}$.

A finite length of the junction or finite temperature both suppress the diode efficiency as can be checked numerically using Eq.~\eqref{eq-current-general}.

 In all of the above results, the limiting factor for the diode efficiency is the Fermi velocity asymmetry. Using typical parameters for InSb~\cite{socInSb} ($E_{so}\sim 5$meV, $\bar \mu\sim m\omega_0^2/2\sim 20$meV), we estimate $\delta v/\bar v$ due to the finite width of the nanowire of the order of $10^{-2}$ far from avoided crossings. By tuning the chemical potential, much larger asymmetries should be achievable. When the asymmetry is induced by a magnetic field, values of $\delta v/\bar v$  of the order of $10^{-1}$ should be reached at fields $B_\perp\sim 1$T taking $g\sim 40$.


 In conclusion,  we presented a simple model for a Josephson diode effect in a single-channel  ballistic nanowire with Rashba spin-orbit coupling and subject to a Zeeman field. The diode effect arises due to the interplay between the Andreev bound states forming within the two pseudo-spin bands. We derived microscopically the current-phase relation that was used phenomenologically in previous works. We showed that the diode effect induced by the magnetic field has two different contributions. One contribution is due to the field in the junction;  
it increases with the length of the junction. Another contribution is due to the field in the leads; it is maximal  in a short junction when only one pseudo-spin band is present.   In general, both contributions may be present. While their relative strength will depend on the details of magnetic screening, the length of the junction is crucial in determining which one is dominant based on the discussion above. The general result, Eq.~\eqref{eq-current-general}, allows on to compute the current-phase relation including both effects. The same physical ingredients also yield an anomalous Josephson current~\cite{JA1,JA2,JA3,multichannel1,multichannel1-prb,JA-recent}. We discuss connections and differences between the two effects in the Supplementary Material.
The experimental challenge is to separate the diode effect due to the interplay between spin-orbit coupling and a Zeeman field from other origins. The predictions given in this work can serve as a guide for assessing the microscopic origin of the diode effect in a concrete setup.

%

\begin{acknowledgments}
We thank Liang Fu and Jean-Damien Pillet for interesting exchanges. Furthermore, we thank Alessandro De Martino for pointing out an error in the estimation of the velocity mismatch. We wish to acknowledge funding from the French
Agence Nationale de la Recherche through Grants No.~ANR-17-PIRE-0001 (HYBRID) and
ANR-21-CE30-0035 (TRIPRES). 
\end{acknowledgments}

\renewcommand{\theequation}{S\arabic{equation}}

\appendix


\section{2D harmonic nanowire with Rashba spin-orbit coupling}
\label{app-NW}

Here we provide a simple twodimensional model of a nanowire with Rashba spin-orbit coupling~\cite{NW-1,NW-last}. The confining potential along the $y$-direction is considered harmonic with frequency $\omega_0$, $V(y)=m\omega_0^2y^2/2$, where $m$ is the effective electron mass. Furthermore, we include an in-plane magnetic field $\vec h=h_x\vec u_x+h_y\vec u_y$. The Hamiltonian takes the form
\begin{eqnarray}
H=\frac{\vec p^2}{2m}+\frac12m\omega_0^2y^2-\alpha(\vec p\times\vec\sigma)\cdot \vec u_z+\vec h\cdot\vec\sigma,
\end{eqnarray}
where $\vec p=(p_x,p_y)$ is the in-plane momentum, $\alpha$ is the strength of the Rashba spin-orbit coupling, $\sigma_i$ ($i=x,y,z$) are Pauli matrices in spin space, and $\vec u_z$ is a unit vector perpendicular to the plane. 

We will consider the transverse spin-orbit coupling ($\sim \alpha p_y$) as well as the field along the wire ($h_x$) as perturbations, namely we will assume $\alpha p_y,h_x\ll\omega_0$. 
The Hamiltonian 
\begin{eqnarray}
H_0=\frac{\vec p^2}{2m}+\frac12m\omega_0^2y^2-\alpha p_x\sigma_y+h_y\sigma_y
\end{eqnarray}
can be easily diagonalized. The eigenfunctions take the form
\begin{eqnarray}
\psi_{k_x,n,\pm}(x,y)=e^{ik_xx}\chi_n(y)\frac1{\sqrt2}\begin{pmatrix}1\\\pm i\end{pmatrix},
\end{eqnarray}
where $\chi_n(y)$ are the harmonic oscillator eigenfunctions, with eigenenergies
\begin{eqnarray}
E_{n\pm}(k_x)=\frac{k_x^2}{2m}+\omega_0(n+\frac12)-\mu\mp(\alpha k_x-h_y).
\end{eqnarray}
Note that the spin states within a subband have the same velocity at a given energy. Band crossings between the two spin states in the first subband take place at $k_x=h_y/\alpha$. Band crossings between opposite spin states in the first and second subband take place at $k_x=\mp\omega_0/(2\alpha)+h_y/\alpha$. The transverse spin-orbit coupling and the longitudinal magnetic field open gaps at these crossings and thus modify the velocities of the different bands and change the spin-projection of the eigenstates.
For sufficiently low chemical potential, we may write an effective Hamiltonian in the space spanned by $\psi_{k_x,0,+},\psi_{k_x,0,-},\psi_{k_x,1,+},\psi_{k_x,1,-}$, namely
\begin{eqnarray}
\tilde H&=&\frac{k_x^2}{2m}-(\alpha k_x-h_y)\sigma_z-\omega_0(1-\frac12\eta_z)\nonumber\\
&&+\eta\sigma_y\eta_y+h_x\sigma_y-\mu
\end{eqnarray}
with $\eta=\alpha\sqrt{m\omega_0/2}\equiv\sqrt{E_{so}\omega_0}$.
The spectrum is given as
\begin{widetext}
\begin{eqnarray}
E_{s_1s_2}=\frac{k_x^2}{2m}+\omega_0-\mu+s_1\sqrt{(\alpha k_x-h_y)^2+h_x^2+\frac{\omega_0^2}4+\eta^2+s_2\sqrt{(\alpha k_x-h_y)^2\omega_0^2+4h_x^2(\frac{\omega_0^2}4+\eta^2)}}.
\end{eqnarray}
Unless the chemical potential is close to an avoided crossing, we may approximate the two lowest bands as
\begin{eqnarray}
E_\pm
&\approx&\frac{(k_x\mp m\alpha)^2}{2m}+\frac{\omega_0}2-\mu-\frac{m\alpha^2}2\mp \left(\frac{h_x^2}{2\alpha k_x}+\frac{\eta^2}{2\alpha k_x\pm\omega_0}\right)\mp h_y\left(1-\frac{h_x^2}{2\alpha^2k_x^2}-\frac{\eta^2}{(2\alpha k_x\pm\omega_0)^2}\right).
\label{eq-spec}
\end{eqnarray}
The Fermi momenta are obtained by setting $E_\pm(k_F^\pm,h_y=0)=0$ whereas the Fermi velocities are obtained as
$$v_\pm=\frac{\partial E_\pm(k_x,h_y=0)}{\partial k_x}|_{k_x=k_F^\pm}=\frac{k_F^\pm}m\mp \alpha\pm \left(\frac{h_x^2}{2\alpha (k_F^\pm)^2}+\frac{2\alpha\eta^2}{(2\alpha k_F^\pm\pm\omega_0)^2}\right).$$
Let us consider the effect of $\eta$ first. At weak spin-orbit coupling, we may develop
\begin{eqnarray}
E_\pm(k_x,h_x=h_y=0)
&\approx&\frac{(k_x\mp m\alpha)^2}{2m}+\frac{\omega_0}2-\mu-\frac{m\alpha^2}2- \frac{\eta^2}{\omega_0}\left(1\mp\frac{2\alpha k_x}{\omega_0}+\left(\frac{2\alpha k_x}{\omega_0}\right)^2\mp\left(\frac{2\alpha k_x}{\omega_0}\right)^3\right).
\end{eqnarray}
\end{widetext}
Only the last term contributes to an asymmetry in the velocities.
Namely
\begin{equation}
v_\pm=\bar v\pm\frac{8\eta^2\alpha^3\bar k_F^2}{\omega_0^4}.
\end{equation}
Here $\bar v=\bar k_F/m$ and $\bar k_F\approx\sqrt{2m(\mu-\frac{\omega_0}2)}$. 
The effective fields are given as
\begin{equation}
h_\pm\approx h_y\left(1\pm\frac{2\eta^2\alpha\bar k_F}{\omega_0^3}\right),
\end{equation}
where we neglected the suppression of $\bar h$. Taking it into account, one recovers $h_\pm\leq h_y$. Note that $\delta v/\bar v\ll\delta h/\bar h$.

We now turn to the effect of $h_x$. With $k_F^\pm\approx\bar k_F\pm m\alpha$, we find
\begin{equation}
v_\pm\approx\bar v\pm\frac{h_x^2}{2\alpha\bar k_F^2}
\end{equation}
and 
\begin{equation}
h_\pm\approx h_y\left(1\pm\frac{mh_x^2}{\alpha\bar k_F^3}\right).
\end{equation}
Here both $\delta v/\bar v,\delta h/\bar h\sim h_x^2m/(\alpha\bar k_F^3)$.

\section{Anomalous Josephson effect}
\label{app-JA}

\setcounter{equation}{12}

The so-called anomalous Josephson effect, i.e., the presence of Josephson current in the absence of an applied phase difference ($\phi=0$), arises from the same interplay between spin-orbit coupling and an external magnetic field. In contrast to the diode effect, a current-phase relation with a single harmonic is sufficient to obtain an anomalous Josephson current.  The anomalous Josephson effect due to the magnetic field in the leads has been studied in detail in Refs.~\onlinecite{dolcini,nesterov}. Here we provide results for the anomalous Josephson current due to the field within the junction.
For a current-phase relation dominated by the first harmonic, it is given as $I_{\rm an}\approx-I_{c1}\sin(\phi_{01})$, except for phases $\phi_{01}\approx n\pi$, where the higher harmonics come into play. At small fields, the anomalous current is linear in $\bar h$ and given as
\begin{equation}
I_{\rm an}\approx\bar I_1\delta\phi+\delta I_1\bar\phi.
\label{eq-an}
\end{equation}
With this, we can compute the anomalous current for a short junction close to $T_c$. It is dominated by the first term in Eq.~\eqref{eq-an} yielding
\begin{equation}
I_{\rm an}^{\rm short}\approx\frac{e\Delta^2}{8T_c}\frac{\bar h}{\bar E_L}\left(\frac{\delta h}{\bar h}-\frac{\delta v}{\bar v}\right).
\end{equation}
Eq.~\eqref{eq-an} also gives the anomalous Josephson current in a long junction at $T\gg \bar E_L$, namely
\begin{equation}
I_{\rm an}^{\rm long}\approx 4eTe^{-2\pi T/\bar E_L}\!\left(\delta\phi\cosh\frac{\pi T\delta v}{\bar E_L\bar v}\!+\!\bar\phi\sinh\frac{\pi T\delta v}{\bar E_L\bar v}\right).\!\!
\end{equation}
 We note that in both cases $|I_{\rm an}|\gg|I_c^>-I_c^<|\propto \bar I_2$.
 
 At $T=0$, the situation is a little more subtle.
 The anomalous current of a short junction is given as
\begin{eqnarray}
I_{\rm an}^{0\,{\rm short}}
&\approx&-\frac{e\Delta}2\sum_{j=\pm}j\sin\frac{\phi_j}2{\rm sign}\left[\cos\frac{\phi_j}2\right].
\end{eqnarray}
To lowest order in field, one obtains $I_{\rm an}\approx-e\Delta\delta\phi/4$. In the narrow interval $\bar\phi-\delta\phi/2<(2n+1)\pi<\bar\phi-\delta\phi/2$, the anomalous current is strongly enhanced reaching values comparable to the critical current, $I_{\rm an}\sim e\Delta$.
In a long junction, he anomalous current to lowest order in the field is given as
 \begin{eqnarray}
 I_{\rm an}^{0\,{\rm long}}
 &=&-e\bar E_L\left(
\frac{\delta\phi}{2 \pi}+\frac{\delta v}{\bar v}\frac{\bar\phi}{2\pi}\right).
\end{eqnarray}
As in the short junction case, in the narrow interval $\bar\phi-\delta\phi/2<(2n+1)\pi<\bar\phi-\delta\phi/2$, the anomalous current is strongly enhanced reaching values comparable to the critical current, $I_{\rm an}\sim e\bar E_L$.

\nocite{*}
\bibliography{aipsamp}

\begin{thebibliography}{99}
\bibitem{note2D-3D}
In a 3D nanowire, typically $a\sim\lambda_F$: in that case, the orbital effect needs to be taken into account for a quantitative prediction of the diode effect at the optimal field.
\bibitem{footnote-gaps}
The magnetic field in the leads may also affect the induced superconductivity, leading to different gaps $\Delta_\pm$ in the two pseudo-spin bands. Here we neglect this effect.
\bibitem{footnote-squid}
Note that the same equations may also be used to  describe an asymmetric SQUID~\cite{squid1,squid2}, where the subscript $j$ labels the two Josephson junctions in the SQUID loop and $\phi_j=j\Phi/(2\Phi_0)$ with $\Phi$ the flux through the loop.
\bibitem{footnote-fu} One obtains the same current-phase relation when finite-momentum pairing is realized in the conventional superconducting leads~\cite{fu-diode,zazunov-diode}.
\bibitem{footnote-topo} Similar results for the case $\Delta_+=\Delta_-$ were obtained in Ref.~\onlinecite{topo}.
\end{thebibliography}

\end{document}